# A Hierarchical Deep Learning Model for Predicting Pedestrian-Level Urban Winds

**Reda Snaiki[1], Jiachen Lu[2,3], Shaopeng Li[4], Negin Nazarian*[2,3,5]**

[1] *Department of Construction Engineering, École de Technologie Supérieure, Université du Québec, Montréal, Québec, Canada*

[2] *School of Built Environment, UNSW Sydney, Sydney, NSW 2052, Australia*

[3] *ARC Centre of Excellence for Climate Extremes, University of New South Wales, Sydney, Australia*

[4] *Department of Civil Engineering and Environmental Engineering, University of Louisiana at Lafayette, LA, USA*

[5] *ARC Centre of Excellence for the 21st Century Weather, UNSW Sydney, Sydney, Australia*

*Corresponding author, n.nazarian@unsw.edu.au

**Abstract:** Deep learning-based surrogate models offer a computationally efficient alternative to high-fidelity computational fluid dynamics (CFD) simulations for predicting urban wind flow. However, conventional approaches usually only yield low-frequency predictions (essentially averaging values from proximate pixels), missing critical high-frequency details such as sharp gradients and peak wind speeds. This study proposes a hierarchical approach for accurately predicting pedestrian-level urban winds, which adopts a two-stage predictor-refiner framework. In the first stage, a U-Net architecture generates a baseline prediction from urban geometry. In the second stage, a conditional Generative Adversarial Network (cGAN) refines this baseline by restoring the missing high-frequency content. The cGAN's generator incorporates a multi-scale architecture with stepwise kernel sizes, enabling simultaneous learning of global flow structures and fine-grained local features. Trained and validated on the UrbanTALES dataset with comprehensive urban configurations, the proposed hierarchical framework significantly outperforms the baseline predictor. With a marked qualitative improvement in resolving high-speed wind jets and complex turbulent wakes as well as wind statistics, the results yield quantitative enhancement in prediction accuracy (e.g., RMSE reduced by 76% for the training set and 60% for the validation set). This work presents an effective and robust methodology for enhancing the prediction fidelity of surrogate models in urban planning, pedestrian comfort assessment, and wind safety analysis. The proposed model will be integrated into an interactive web platform as Feilian Version 2.

**Keywords:** Urban wind flow; Urban climate modeling; Image-to-image translation; Hierarchical model; Deep learning.

## 1. Introduction

The accurate characterization of wind flow within the urban canopy is a longstanding challenge in environmental engineering and city planning, with direct implications for pedestrian comfort, pollutant dispersion, and the structural safety of buildings and urban infrastructures (Blocken, 2015; Mittal et al., 2018). Historically, three primary methodologies have been employed to study this complex phenomenon: in-situ measurements, wind tunnel and reduced-scale experiments, and computational fluid dynamics (CFD) simulations (Grimmond et al., 1998; Zhong et al., 2022; Nazarian et al., 2023). In-situ measurements, which rely on data from fixed sensors such as anemometers, ultrasonic transducers, and local weather stations, provide highly accurate ground-truth data (Fenner et al., 2024; Lyu et al., 2025). However, these point-based measurements are often too sparse to resolve the intricate spatial variability of urban wind. Wind tunnel experiments offer a controlled environment for studying flow around scaled models but can be subject to scaling uncertainties and limitations in replicating real-world atmospheric conditions (Ng et al., 2011; Shen et al., 2020; Chen et al., 2022; Li et al., 2024). In recent decades, CFD simulations, particularly high-fidelity methods like Large-eddy Simulation (LES), have become the state-of-the-art for generating comprehensive, three-dimensional datasets of urban wind flow (Blocken, 2015; Buccolieri et al., 2021; Lu et al., 2023a and 2024). Despite their accuracy, the immense computational cost of these physics-based simulations, which often require hours or days of high-performance computing for a single scenario, renders them impractical for applications requiring rapid design iteration or large-scale analysis.

The computational bottleneck of CFD has motivated the development of machine learning-based surrogate models that emulate the input-output mapping of physics-based solvers at a fraction of the cost (Brunton et al., 2020; Xie et al., 2020; Wu and Snaiki, 2022). Trained on large datasets generated from high-fidelity CFD simulations, these models have evolved rapidly, progressing from foundational methods like k-Nearest Neighbors (kNN) (BenMoshe et al., 2023) to more complex deep learning frameworks (Xiao et al., 2019; Xie et al., 2020; Kastner and Dogan, 2023; Clarke et al., 2025). Convolutional Neural Networks (CNNs), particularly the U-Net architecture, have since become standard for framing the task as an image-to-image translation problem, mapping urban geometry to wind field predictions (Lu et al., 2023b; Vargiemezis and Gorlé, 2025). More advanced paradigms include Fourier Neural Operators (FNOs), which learn resolution-invariant operators in the frequency domain for greater efficiency and generalizability (Peng et al., 2024), and Physics-Informed Neural Networks (PINNs) (Gråberg, 2022), which embed the governing Navier-Stokes equations directly into the learning process to enforce physical consistency and reduce reliance on large datasets. Furthermore, Graph Neural Networks (GNNs) are an emerging approach well-suited to the irregular topology of urban environments, having demonstrated performance comparable to CNNs in related forecasting tasks (Yu et al., 2024). These advances have enabled prediction speeds several orders of magnitude faster than traditional solvers while retaining strong agreement with CFD-generated ground truth. Nonetheless, a persistent limitation remains (Lu et al., 2025): pixel-wise regression models tend to produce overly smooth predictions, compressing the wind speed distribution to its mode over idealized urban arrays and even erasing local extrema over realistic urban neighborhoods, which are all critical for assessing wind extremes in the built environment.

To address this research gap, this paper introduces a novel two-stage, hierarchical predictor-refiner framework that significantly enhances the fidelity of surrogate model predictions. The first stage employs a U-Net-based predictor to generate a stable, low-frequency baseline from urban geometry. The second stage applies a conditional Generative Adversarial Network (cGAN) as a refiner to restore high-frequency details absent in the baseline. The cGAN's generator integrates a multi-scale architecture with stepwise kernel sizes, enabling it to capture both large-scale flow structures and small-scale turbulent features and wind extremes. The proposed two-stage hierarchical model was trained and validated on the high-resolution UrbanTALES dataset (Nazarian et al. 2025) that covers neutral simulations over both idealized building arrays and realistic urban neighborhoods to assess its refinement relative to the original U-Net model (Lu et al., 2025).The proposed model, named Feilian Version 2, will be integrated into an interactive web platform as the improved version of U-Net-based Feilian Version 1 (Note that Feilian is the name of a Chinese wind spirit).

## 2. Methodology

This section details the proposed methodology for enhancing surrogate model predictions of urban wind flow. The goal is to develop a hierarchical deep learning framework capable of accurately and efficiently predicting the 2D pedestrian-level wind speed field (at a height of 1.5 meters). This framework is designed to first generate a robust baseline prediction (Lu et al., 2025) and subsequently refine it using a generative adversarial network to emulate the high-fidelity detail of LES. The following subsections describe the data source to define the problem, the overall two-stage framework, the architecture of each component, and the final training protocol.

## 2.1. Data source and problem definition

The dataset used to train and validate the proposed framework is UrbanTALES, an extensive collection of 512 high-resolution Large-eddy Simulations (LES) of pedestrian-level wind. These simulations provide the high-fidelity ground-truth data for the surrogate modeling task. The dataset includes cases ranging from idealized building arrays to realistic neighborhoods extracted from OpenStreetMap (2017), spanning diverse urban densities, height distributions, and wind orientations.

The problem is framed as an image-to-image translation task. The primary input for the overall framework is the urban geometry, represented as a 2D map of building heights for a given scenario and a prevailing wind direction. The desired output is the corresponding 2D map of the time-averaged wind speed at a pedestrian-relevant height of 1.5 meters, as provided by the LES ground-truth fields. The objective is to learn this mapping from geometry to the high-fidelity wind field with high computational efficiency and accuracy.

## 2.2. Overall framework

The proposed methodology is founded on the principle of decomposing the complex, direct mapping from urban geometry to a high-fidelity wind field into two distinct, more manageable sub-tasks, an approach conceptually similar to hierarchical or cascaded learning methods (e.g., Viola and Jones, 2001; Silla and Freitas, 2011). This two-stage "predictor-refiner" framework, illustrated in Fig. 1, is designed to improve both the stability of the training process and the quality of the final prediction.

The first stage, the predictor, utilizes a pre-trained U-Net model based on the architecture of Lu et al. (2025). This model serves to establish a computationally efficient, physically plausible baseline prediction. Its primary role is to learn the low-frequency components of the flow field, effectively capturing the large-scale flow structures and dominant wind paths dictated by the urban morphology. The second stage, the refiner, introduces a conditional Generative Adversarial Network (cGAN). This network is not trained on the urban geometry directly; instead, its sole input is the baseline prediction from Stage 1, which is tasked to transform it into a high-fidelity output. The refiner learns to add the missing high-frequency details, such as sharp velocity gradients and peak wind speeds, thereby correcting the systematic smoothing artifacts inherent in the baseline model. By constraining the cGAN to this refinement task, the training process is stabilized and the network's capacity is focused entirely on learning these high-frequency components of the flow.

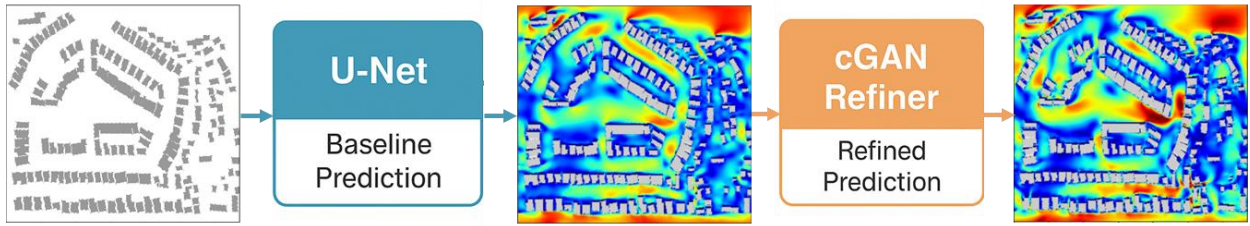

**Fig. 1** Overview of the two-stage predictor-refiner framework. Stage 1: A U-Net baseline predictor maps urban geometry (2D array of building height shown by the grayscale color) and inflow conditions to a rough speed field. Stage 2: A conditional GAN refiner transforms the rough prediction into a high-fidelity output matching LES-level detail.

**2.3. Stage 1: Baseline predictor model**

The first stage of the proposed framework is the predictor, which is responsible for generating a rapid, baseline approximation of the wind field. For this role, the pre-trained U-Net surrogate model developed by Lu et al. (2025) is employed. This model was selected due to its validated ability to efficiently produce physically plausible 2D wind speed maps directly from urban geometry.

The predictor model functions as a direct mapping from the urban form to the resulting pedestrian-level wind field. The input consists of a 2D array representing urban building heights, coupled with the prevailing wind direction. The model's output is a 2D map of the time-averaged wind speed at a pedestrian-relevant height of 1.5 meters. To handle varied urban layouts and wind angles, the model utilizes a sophisticated pre-processing scheme where, for non-orthogonal winds, the urban geometry is rotated to align the flow with the image axes. This canonization of the input simplifies the learning task for the network (see Lu et al. 2025 for more information on the methodology).

The employed model is a U-Net, a type of fully convolutional neural network (CNN) particularly effective for image-to-image translation tasks, as illustrated in Fig. 2. Lu et al. (2025) adapted the standard U-Net architecture for this specific physical problem by incorporating additional downsampling layers for hierarchical feature extraction, batch normalization for training stability, and a final ReLU activation function to ensure the physically necessary constraint of non-negative wind speed outputs. The model was trained on the comprehensive UrbanTALES dataset using a Mean Absolute Error (MAE) loss function. In the context of this work, this pre-trained (based on the entire UrbanTALES dataset) and validated model is used directly to generate the baseline predictions that serve as the input for the subsequent generative refiner network.

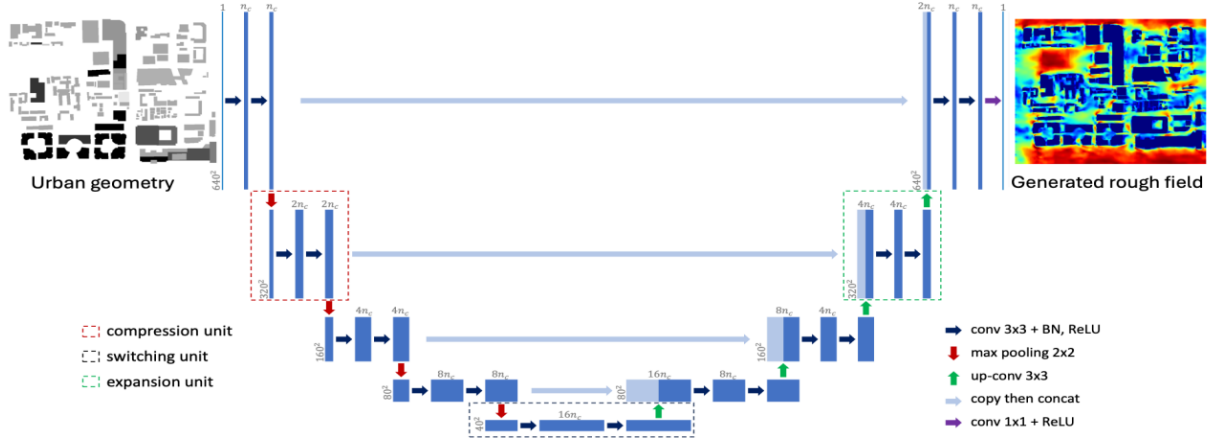

**Fig. 2** Schematic of the Stage-1 U-Net baseline predictor, showing the encoder–decoder structure and skip connections used for rough estimation, adapted from Lu et al., (2025).

### 2.4. Stage 2: Generative adversarial refiner network

The second stage of the framework employs a conditional Generative Adversarial Network (cGAN) to perform the critical refinement task. This network is designed to execute a sophisticated image-to-image translation, transforming the low-fidelity baseline prediction into a high-fidelity output that emulates the ground truth. The cGAN framework consists of two neural networks, a Generator and a Discriminator, which are trained simultaneously in an adversarial process (Goodfellow et al., 2014; Mirza, M. and Osindero et al., 2014). The proposed implementation is based on the Pix2Pix architecture (Isola et al., 2017), which is highly effective for paired image translation problems.

The overall architecture of the proposed cGAN refiner is illustrated in Fig. 3. The U-Net-based generator takes as input the baseline wind field predicted by Stage 1 and outputs a refined, high-fidelity wind field. Both the generated and ground-truth wind fields are concatenated with the source image and passed to the PatchGAN discriminator (Demir and Unal, 2018), which classifies local patches as 'real' or 'fake'. The generator is trained using a combination of pixel-wise reconstruction loss (L1 loss) and adversarial loss, enabling it to produce outputs that are both quantitatively accurate and perceptually realistic (Johnson et al., 2016; Isola et al., 2017; Yousif et al., 2022).

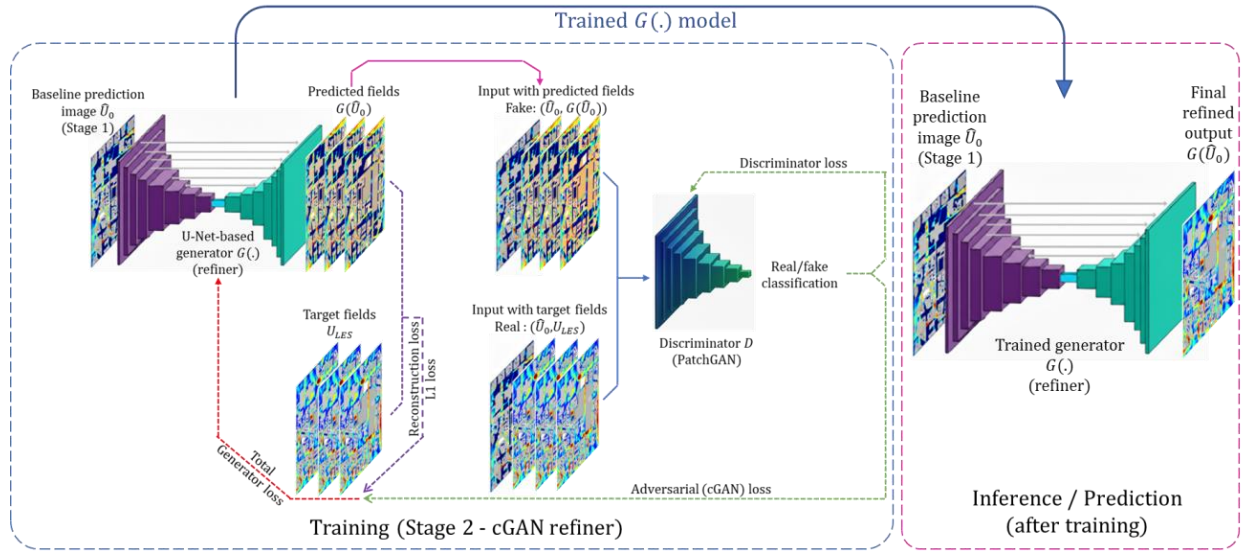

**Fig. 3** Overall architecture of the proposed cGAN refiner, showing the generator, discriminator, and loss computation pathways.

### 2.4.1. Enhanced generator architecture

The generator is responsible for producing the final refined wind speed map from the baseline prediction. To achieve the required level of detail and accuracy, a standard U-Net is insufficient. Therefore, an enhanced generator architecture was developed, based on a U-Net structure but with significant modifications to improve its capacity for multi-scale feature learning. The overall role of the generator within the cGAN framework is shown in Fig. 3.

The core of the architecture is the incorporation of stepwise kernel sizes within an 8-layer encoder. Unlike traditional U-Nets that use a small, fixed-size kernel throughout, this design employs large kernels in the upper encoder to capture broad, low-frequency spatial relationships, and progressively smaller kernels in deeper layers to extract high-frequency details. Concretely, the encoder uses strided convolutions (stride 2) with kernel sizes 32, 32, 16, 16, 8, 8, 8, and 4 across successive levels, with instance normalization in all encoder blocks except the first and LeakyReLU ($\alpha = 0.2$) activations (Xu et al., 2020). Channel width increases from 64 up to 512 and is capped at 512 in the deepest blocks, consistent with common U-Net practice for stable capacity scaling.

The decoder mirrors the encoder structure and restores resolution using transpose convolutions (kernel 4, stride 2, padding 1). Each decoder block applies instance normalization and ReLU activations, and skip connections concatenate encoder features at matching resolutions to preserve high-resolution spatial information critical for precise reconstruction. To mitigate overfitting while the network learns fine-scale corrections, dropout (p=0.5) is applied in the three deepest decoder stages. The final layer is a transpose convolution followed by a Tanh activation, mapping the single-channel output to the normalized $[-1,1]$ range required for stable GAN training and consistent with the data normalization used for targets.

### 2.4.2. Discriminator architecture

The discriminator drives the adversarial training process by distinguishing between refined predictions produced by the generator and real LES ground-truth data. Within the framework shown in Fig. 3, it receives as input a two-channel tensor created by concatenating the source image (the baseline U-Net prediction) with either the real target image (LES truth) or the generator's refined prediction.

A PatchGAN architecture is used, consistent with the Pix2Pix framework. Rather than outputting a single scalar 'real' or 'fake' score for the entire image, the PatchGAN operates on overlapping image patches, classifying each one individually. This approach is computationally efficient and encourages the generator to produce realistic high-frequency textures and sharp details throughout the output. Concretely, the discriminator is implemented as a deep CNN composed of successive 4×4 strided convolutions (stride 2, padding 1) with instance normalization and LeakyReLU ($\alpha$=0.2) activations; the first block omits normalization for stability. A final 4×4 convolution (stride 1, padding 1) produces a 2-D logit map (no sigmoid), where each element scores the realism of a corresponding patch in the input; these logits are subsequently consumed by a BCE-with-logits objective.

### 2.5. Training objective and protocol

### 2.5.1. Training objective

Training the cGAN refiner follows the paired image-translation paradigm of Pix2Pix and uses a composite loss that combines an adversarial term (to encourage perceptual realism) with a pixel-wise reconstruction term (to preserve quantitative fidelity). Let $x$ denote the Stage-1 baseline prediction (the generator input), $y$ the LES ground truth, and $G(x)$ the generator output. Let $D$ denote the PatchGAN discriminator operating on paired inputs; given a source $x$ and a target $y$ (or the generated output $G(x)$), it produces a pre-sigmoid logit map $f(x, y)$ with one logit per patch, and the corresponding probability map is $D(x, y) = \sigma(f(x, y))$, where $\sigma$ is the elementwise sigmoid. All expectations below are averaged over the training minibatch and over patches. The conditional adversarial objective is:

$$\mathcal{L}_{cGAN}(G, D) = \mathbb{E}_{x,y}[\log D(x, y)] + \mathbb{E}_x[\log(1 - D(x, G(x)))] \tag{1}$$

which the discriminator $D$ maximizes and the generator $G$ minimizes. In practice, this is implemented using a binary cross-entropy (BCE) loss. Since the PatchGAN discriminator outputs a map of predictions for each patch, the final loss is calculated by averaging the BCE loss across all patches.

To ensure that the refined field remains quantitatively close to the LES target, a pixel-wise reconstruction loss is added. Following Pix2Pix, the L1 norm is used, which is empirically found to encourage less blurring than L2:

$$\mathcal{L}_{L1}(G) = \mathbb{E}_{x,y}[\|y - G(x)\|_1] \tag{2}$$

The generator's total loss is a weighted sum of the adversarial loss (implemented as the negative log probability of the discriminator labeling the generated examples as real) and the L1 reconstruction term:

$$\mathcal{L}_G = \mathbb{E}_x[-\log D(x, G(x))] + \lambda \mathcal{L}_{L1}(G) \tag{3}$$

Consistent with Pix2Pix, $\lambda = 100$ is selected to strongly favor faithful reconstruction while still encouraging realistic high-frequency detail. Equivalently, optimization alternates (or concurrently updates) to solve:

$$\min_G \max_D \mathcal{L}_{cGAN}(G, D) + \lambda \mathcal{L}_{L1}(G) \tag{4}$$

The discriminator loss (implemented with BCE over real and fake patches) is:

$$\mathcal{L}_D = -\mathbb{E}_{x,y}[\log D(x, y)] - \mathbb{E}_x[\log(1 - D(x, G(x)))] \tag{5}$$

i.e., the negative of $\mathcal{L}_{cGAN}$ so that minimizing $\mathcal{L}_D$ drives $D$ to classify real pairs as 1 and generated pairs as 0.

### 2.5.2. Training protocol

The refiner network requires paired examples of baseline predictions (from stage 1) and corresponding LES ground truth to learn the refinement task. Any NaN values corresponding to building locations in the raw data arrays were replaced by zeros prior to resizing and normalization. Each paired example was resized to a uniform spatial dimension of 512×512 pixels to enable efficient batch training. Pixel values for both baseline inputs and LES targets were linearly scaled to the range $[-1,1]$ prior to training (the generator uses a final tanh activation function). Crucially, the scaling factors (per-channel maxima) were computed on the training split only and then applied unchanged to the validation split to avoid leakage; diagnostic images saved during training were re-scaled back to physical units using these training-derived maxima. After resizing, the baseline channel was clamped to non-negative values to maintain consistency with the Stage-1 predictor's range. To assess generalization, the available paired dataset was split into a training set (90%) and a validation set (10%). The refiner was trained exclusively on the training split, while the validation split was used only for checkpointing and early stopping.

The model was trained following the paired image-translation paradigm described in Section 2.5.1. The following settings were used: a generator learning rate of $2 \times 10^{-4}$, a discriminator learning rate of $1 \times 10^{-4}$, the Adam optimizer, and a batch size of 4 for a maximum of 500 epochs. The L1 weight $\lambda$ was set to 100. The adversarial term was computed using BCEWithLogitsLoss for numerical stability. When updating the discriminator, generated samples were detached from the generator's computation graph to prevent backpropagation into $G$. Instance normalization was used in the majority of convolutional blocks to stabilize training, and dropout was applied in the deepest decoder stages to reduce overfitting. An early stopping protocol with a patience of 20 epochs on the validation L1 loss was employed; the checkpoint achieving the minimum validation L1 was retained for all subsequent analyses.

## 3. Results and Discussion

This section presents a comprehensive evaluation of the proposed two-stage predictor-refiner framework. The model's performance is assessed across a range of test cases, including diverse, real-world urban configurations and simplified, idealized geometries. The robustness of the model to variations in the inflow wind direction is also investigated. The evaluation combines qualitative visual comparisons against the LES ground truth with quantitative analysis using standard error metrics.

### 3.1. Training dynamics and model convergence

The cGAN refiner was trained following the Pix2Pix protocol described in Sec. 2.5.1, with a BCE-with-logits adversarial objective and an L1 reconstruction term weighted by $\lambda_{L1} = 100$. Inputs and targets were resized to 512×512 and linearly scaled to $[-1,1]$ using statistics from the training split. For the supervised target channel, the training-set maximum was $max_B = 2.7932\ ms^{-1}$. Optimization used Adam for the generator and discriminator with learning rates $2 \times 10^{-4}$ and $1 \times 10^{-4}$, respectively, batch size 4, a budget of 500 epochs, checkpointing on validation L1, and early stopping with a patience of 20 epochs.

Validation error decreased steadily through mid-training and then entered the characteristic oscillatory regime of adversarial learning. The best checkpoint was obtained at epoch 146 with a validation L1 of 0.029 (normalized), which corresponds to a mean absolute error of approximately $0.041\ ms^{-1}$. At this epoch, the training L1 was 0.007 ($\approx 0.0103\ ms^{-1}$), yielding a generalization gap of 0.022 ($\approx 0.031\ ms^{-1}$). Training subsequently continued without surpassing this minimum and terminated at epoch 166 due to early stopping. In a $\pm 10$-epoch window around the best epoch, the standard deviation of the validation L1 was 0.004 ($\approx 0.0056\ ms^{-1}$), indicating a flat and stable minimum without late-epoch drift.

Adversarial dynamics were characterized by discriminator loss statistics across epochs and by decomposition of the generator objective at the selected checkpoint. The discriminator loss at epoch 146 was 0.6749, close to $\ln 2 \approx 0.693$, the theoretical value when real and generated samples are comparably hard to distinguish. Over the full trajectory, the median discriminator loss was 0.3248 with an interquartile range of 0.5494, reflecting intermittent phases of discriminator strength interleaved with balanced play. Measured as epoch fractions, 44.6% of epochs lay in a balanced band (discriminator loss 0.4–1.0), 36.7% showed discriminator saturation (loss < 0.2), and only 1.2% indicated a weak discriminator (loss > 1.2). Decomposing the generator objective at epoch 146 yields an adversarial term $G_{adv} = 0.8658$ and a GAN-to-reconstruction ratio of 1.179, indicating that both the pixel term and the adversarial signal contributed meaningfully at selection time, consistent with sharper flow details without sacrificing pixel fidelity. Figure 4 reports the per-epoch training and validation L1 curves. All downstream quantitative and qualitative evaluations use this best-validation checkpoint (epoch 146).

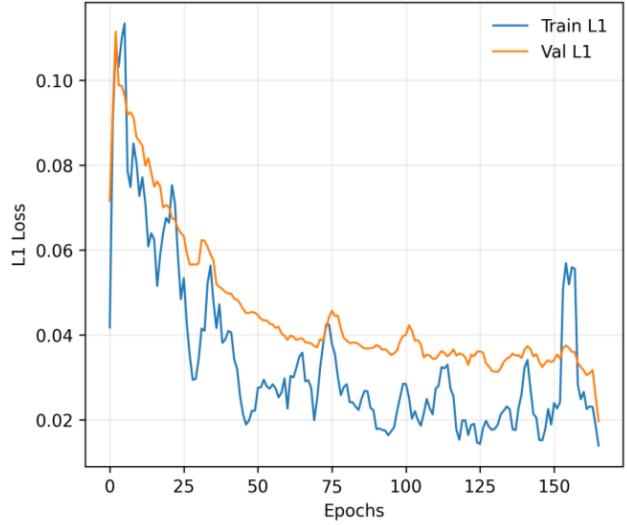

**Fig. 4** Training and validation L1 versus epoch

Beyond the training-curve and convergence diagnostics (Fig. 4), generalization is evaluated at dataset scale. The best-validation checkpoint is applied to all available cases (training and validation). Figures 5–7 present distributions of per-case $R^2$, RMSE, and MAE comparing the Baseline U-Net and the Hierarchical Model for (i) all cases, (ii) real configurations, and (iii) idealized configurations. For readability, panels reporting $R^2$ exclude cases with negative baseline $R^2$. The evaluation follows the same normalization and protocol described in Sec. 2.

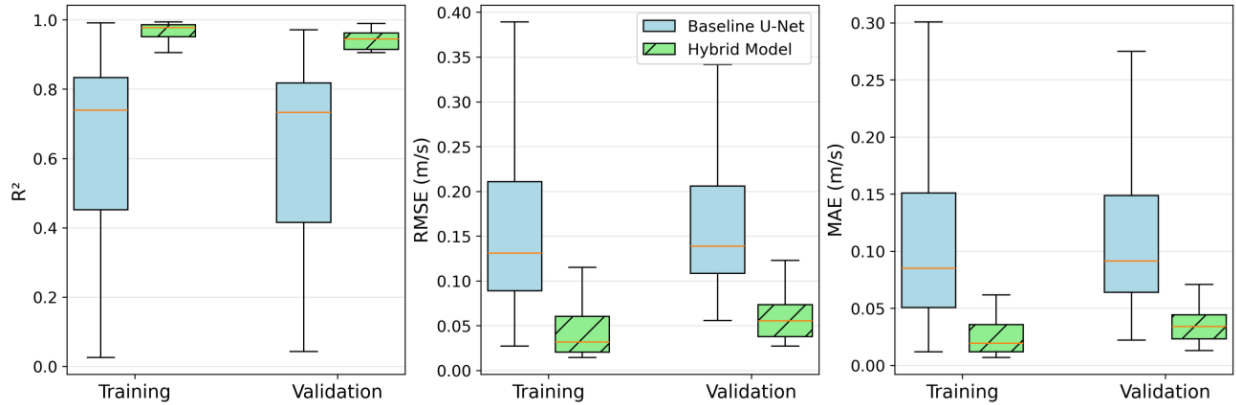

**Fig. 5** Comparison of model performance on the complete dataset. The box plots illustrate the per-case distribution of (a) R² (left), (b) RMSE (middle), and (c) MAE (right) for the Baseline U-Net and the Hierarchical Model, benchmarked against the LES ground truth data.

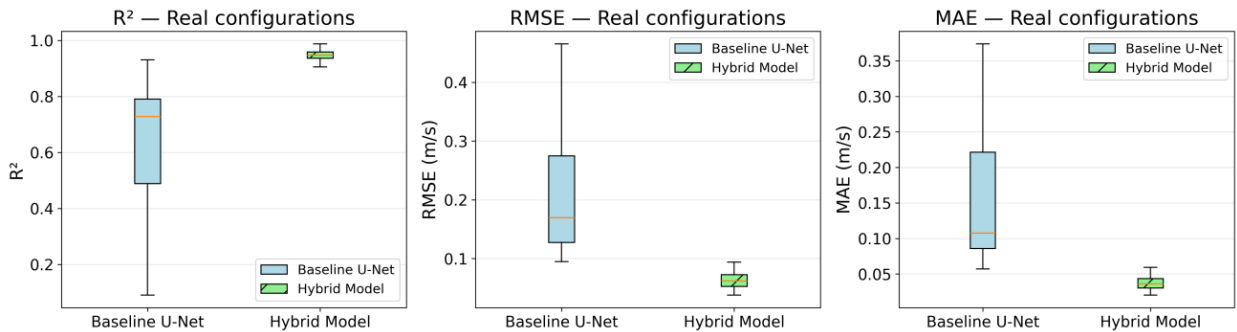

**Fig. 6** Performance evaluation on real urban configurations. The box plots present the same comparison as in Fig. 5, showing per-case distributions of (a) R² (left), (b) RMSE (middle), and (c) MAE (right) but restricted to the subset of realistic urban environments.

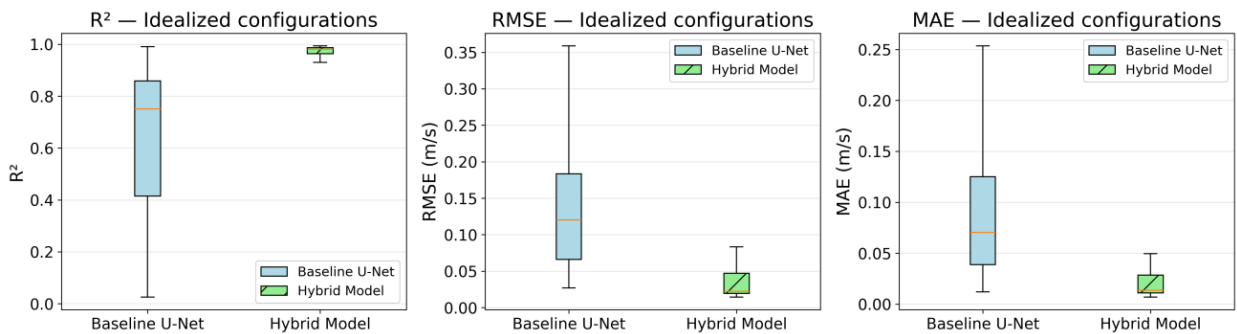

**Fig. 7** Performance evaluation on idealized configurations. The box plots present the same comparison as in Fig. 5, showing per-case distributions of (a) R² (left), (b) RMSE (middle), and (c) MAE (right) but restricted to the subset of idealized geometric layouts.

Using the best-validation checkpoint, the Hierarchical Model outperforms the Baseline U-Net on both data splits and across configuration types. For the training split (Fig. 5), median RMSE decreases from 0.131 to 0.032, a 75.6% reduction; median MAE decreases from 0.085 to 0.019, a 77.6% reduction; median R² increases from 0.573 to 0.974, a 69.9% relative increase. For the validation split (Fig. 5), median RMSE decreases from 0.139 to 0.056, a 59.7% reduction; median MAE decreases from 0.091 to 0.034, a 62.6% reduction; median R² increases from 0.573 to 0.945, a 64.9% relative increase. By configuration type, real urban cases (Fig. 6) show MAE decreasing from 0.108 to 0.036 (66.7% reduction), RMSE from 0.169 to 0.062 (63.3% reduction), and R² increasing from 0.537 to 0.943, a 75.6% relative increase; idealized layouts (Fig. 7) show MAE decreasing from 0.070 to 0.014 (80.0% reduction), RMSE from 0.120 to 0.022 (81.7% reduction), and R² increasing from 0.589 to 0.982, a 66.7% relative increase. Overall, the results indicate robust generalization and systematic error reduction under a single checkpoint and evaluation protocol.

The hierarchical model also demonstrates significant improvements in dataset-level wind-field statistics over the Baseline U-Net. On the training split, the MAE for the spatial mean dropped by 68.1% (from 0.0166 m/s for the baseline to 0.0053 m/s for the hierarchical), while the MAE for spatial standard deviation fell by 79.2% (from 0.0178 m/s to 0.0037 m/s). The error for maximum wind speed was also reduced by 35.5%. These improvements show strong generalization to the validation split, which saw MAE reductions of 41.5% for the mean, 57.7% for the standard deviation, and 32.8% for the maximum wind speed. Taken together with the case-wise error metrics, these dataset-level statistics indicate that the refiner consistently reduces bias in the bulk flow (mean), tightens variability estimates (std), and curbs extremes (max) across both splits.

### 3.2. Case studies: real urban configurations

The performance of the predictor-refiner framework is illustrated on a set of six realistic city geometries from the UrbanTALES dataset (Nazarian et al. 2025) that cover 1) a compact ($\lambda_p = 0.49$) commercial core district, 2) a sparse ($\lambda_p = 0.21$) suburban residential area, 3) a compact ($\lambda_p = 0.47$) modern residential grid, 4) a compact ($\lambda_p = 0.46$) dense residential blocks, 5) an intermediate ($\lambda_p = 0.28$) mixed-use urban blocks, and 6) a sparse ($\lambda_p = 0.14$) modern residential grid where the original Lu et al. (2025) model performs poorly. The evaluation combines a qualitative visual analysis of the generated wind fields with a quantitative assessment using standard error metrics. Figures 8 and 9 present a visual comparison of the model's performance for the six different selected city configurations under a 0° prevailing wind. In all cases, the baseline predictor (left panel in each figure) captures the broad spatial distribution of the wind, identifying the primary wind corridors and low-speed zones. However, the predictions are characterized by excessive smoothing and a failure to resolve localized, high-speed wind jets. In contrast, the refined predictions (middle panel) demonstrate a marked increase in fidelity. Across all different six cities, the refiner network successfully sharpens the flow field. It restores sharp gradients and adds back critical details in wake regions, producing a mean flow that closely match the LES ground truth (left panel).

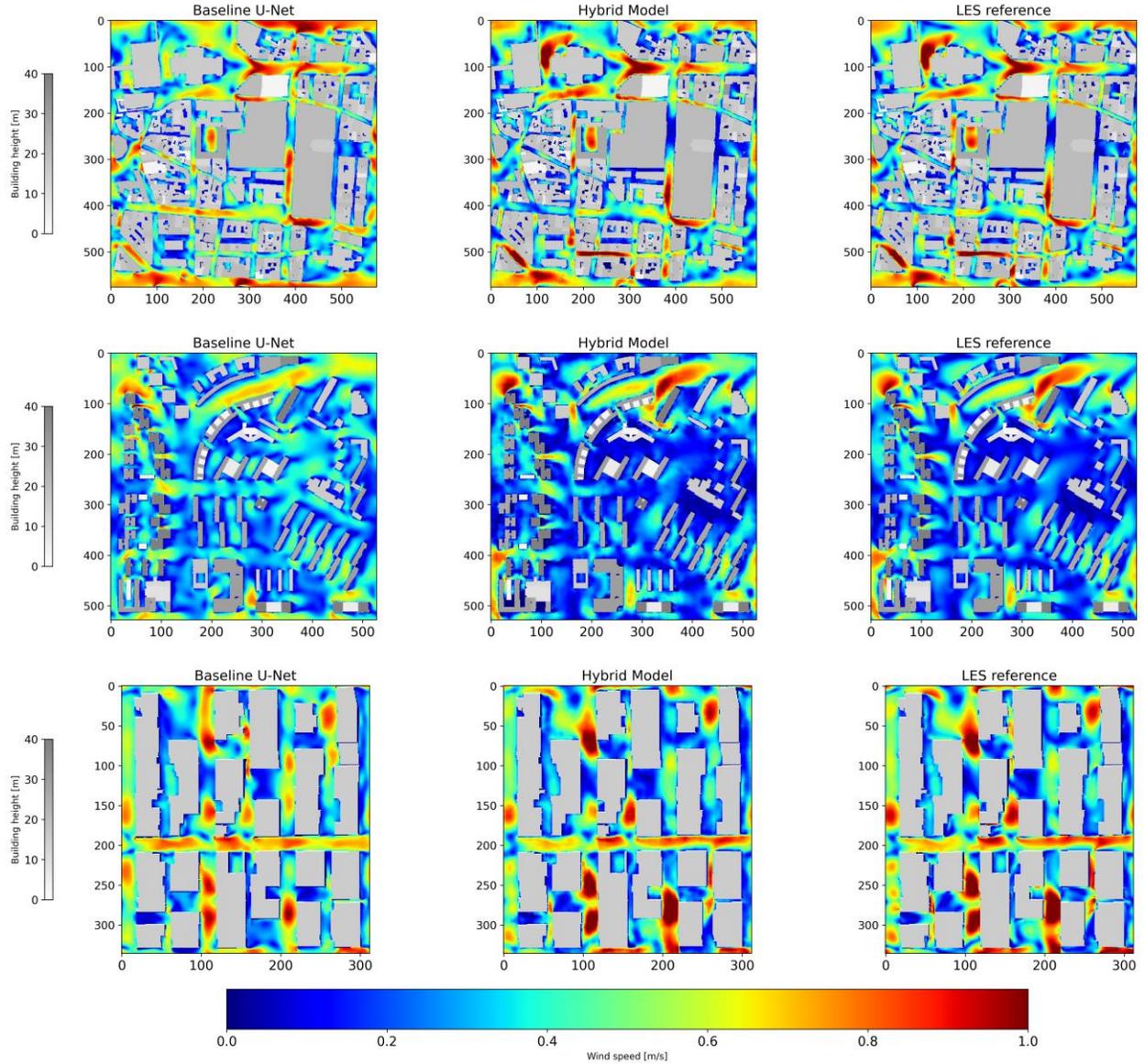

**Fig. 8** Qualitative comparison (0° wind). Each row presents a different city: (top) FR-PA-V2_d00, (middle) ES-Bar-V1_d00, and (bottom) CA-Van-U2_d00. Columns compare the baseline U-Net prediction (left), the hierarchical model prediction (middle), and the LES ground truth (right). The naming convention can be found in Nazarian et al. (2025)

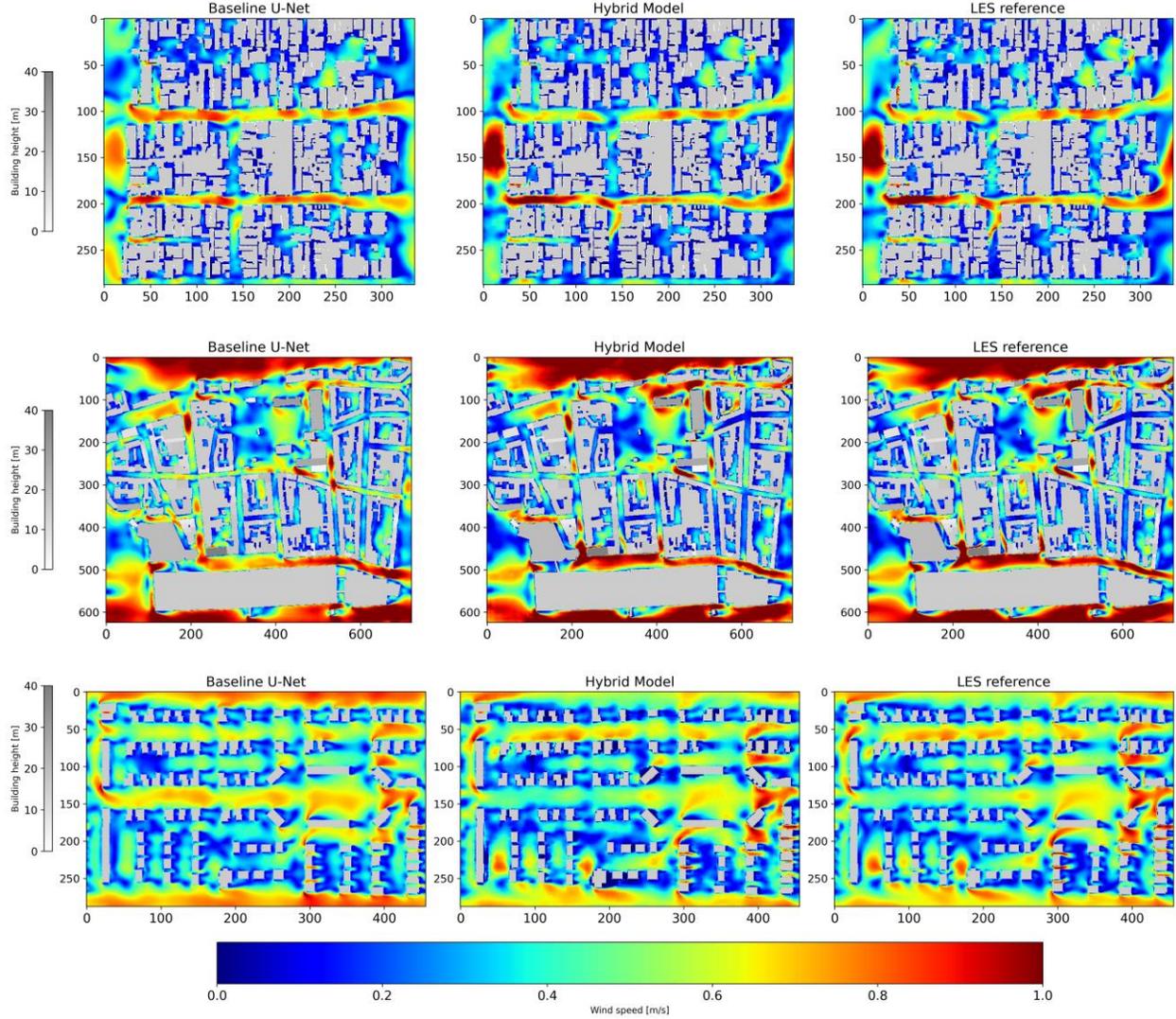

**Fig. 9** Qualitative comparison (0° wind). Each row presents a different city: (top) BR-Sao-U2_d00, (middle) CH-Bas-V1_d00, and (bottom) US-Det-U9_d00. Columns compare the baseline U-Net prediction (left), the hybrid model prediction (middle), and the LES ground truth (right). The naming convention can be found in Nazarian et al. (2025)

The visual improvements are corroborated by Fig. 10. Across the six test cities, mean RMSE decreases by 67% and median RMSE decreases by 60%. Per-city RMSE decreases are FR-PA-V2: 66.0%, ES-Bar-V1: 66.3%, CA-Van-U2: 59.7%, BR-SP-U2: 81.6%, CH-Bas-V1: 54.7%, and US-Det-U9: 50.7%. Variability also tightens markedly, with the RMSE standard deviation reduced by ≈90%. For $R^2$, the mean increases by ≈141% and the median increases by ≈130%; moreover, the share of cases with $R^2 \geq 0.90$ rises from 0% to 100%. Dispersion is similarly reduced, with the $R^2$ standard deviation dropping by ≈95%. Collectively, these percentage gains indicate strong generalization across morphologically diverse cities and reliable recovery of the fine-scale structure required for accurate pedestrian-level wind estimation.

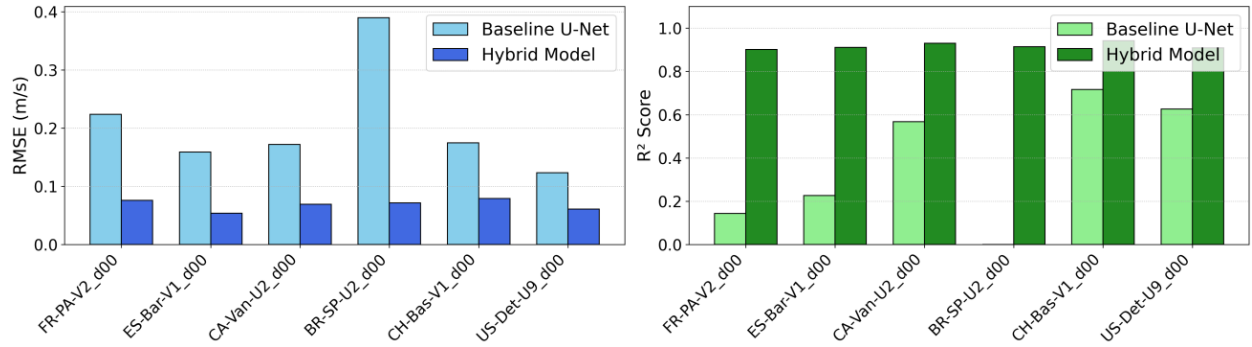

**Fig. 10** Quantitative comparison of model performance across six diverse city configurations. The two subplots compare the performance of the baseline U-Net against the final refined model. (Left) Root Mean Square Error (RMSE); (Right) R² Score.

### 3.3. Case studies: idealized configurations

The capability of the proposed model on regular geometries was assessed using three idealized building-array cases: UA53_d00, US16_d00, and VA16-CL-28_d00 as illustrated in Fig. 11. The Baseline U-Net captures the dominant channelized flow through the arrays but retains a smoothed appearance that blurs shear layers and wake recirculation. The Hierarchical Model (predictor + refiner) sharpens these features, reinstating concentrated high-speed streaks along building edges and clarifying wake decay patterns visible in the LES reference.

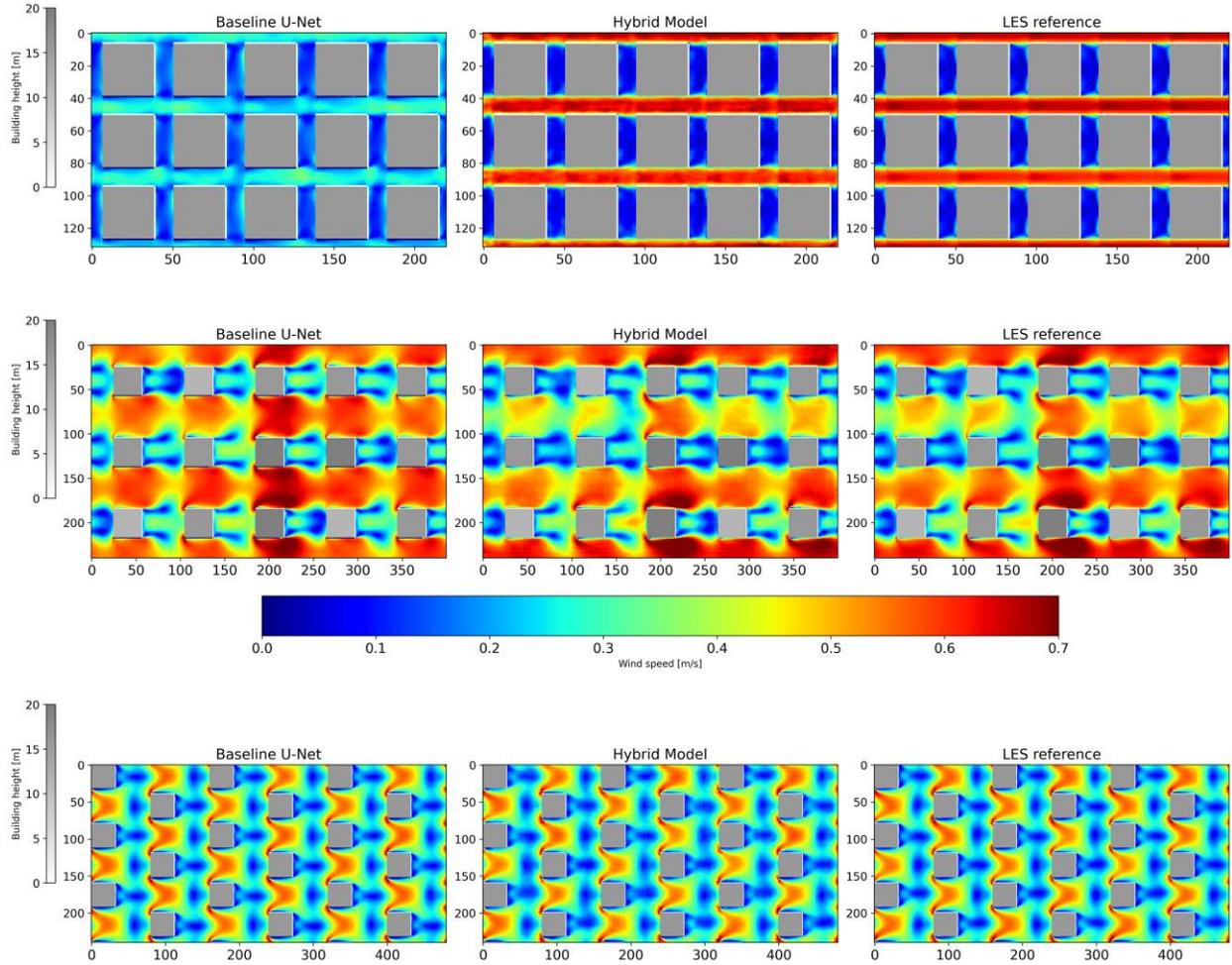

**Fig. 11**. Qualitative comparison for idealized urban geometries. Each row shows a different idealized test case under a 0° wind direction, comparing the baseline U-Net prediction (left), the final refined prediction (middle), and the LES ground truth (right). Cases shown: UA53_d00, US16_d00, and VA16-CL-28_d00.

For quantitative context, Fig. 12 summarizes the per-case RMSE and $R^2$ metrics for both models. The error magnitudes decrease substantially after refinement. For instance, the RMSE for case UA53_d00 was reduced by 90.5% (from 0.3638 to 0.0346 m s$^{-1}$), while cases US16_d00 and VA16-CL-28_d00 saw reductions of 83.2% (from 0.1363 to 0.0229 m s$^{-1}$) and 83.1% (from 0.1441 to 0.0243 m s$^{-1}$), respectively. Across these three cases, the mean and median RMSE reductions were 85.6% and 83.2%. The goodness-of-fit also improved significantly, with $R^2$ values increasing

from -1.135 to 0.981 (UA53_d00), 0.050 to 0.973 (US16_d00), and 0.254 to 0.979 (VA16-CL-28_d00).

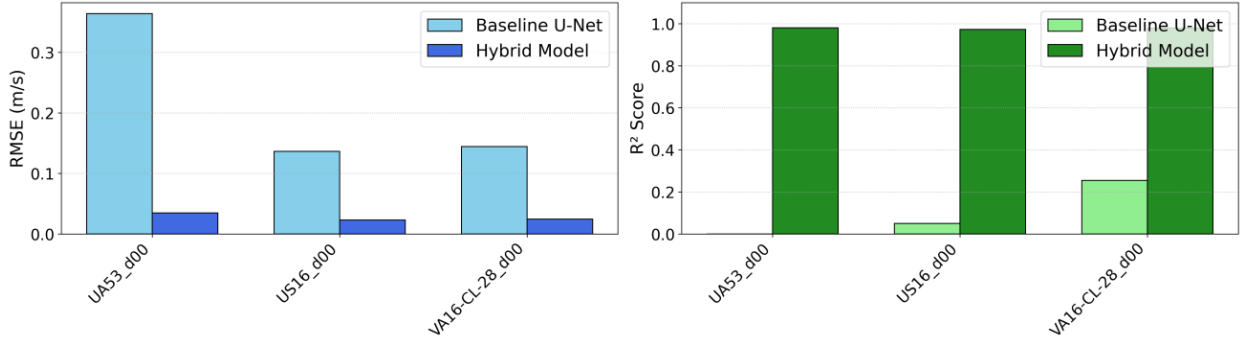

**Fig. 12** Quantitative comparison of model performance across three diverse idealized configurations. The two subplots compare the performance of the baseline U-Net against the final refined model. (Left) Root Mean Square Error (RMSE); (Right) R² Score.

### 3.4. The effect of wind direction

A key test of a surrogate model's robustness is its ability to generalize to different inflow directions, as this significantly alters the complex aerodynamic interactions within the urban canopy. To evaluate the framework's directional robustness, the KO-SE-V11 test case was evaluated for five inflow angles: 0°, 15°, 30°, 45°, and 90°.

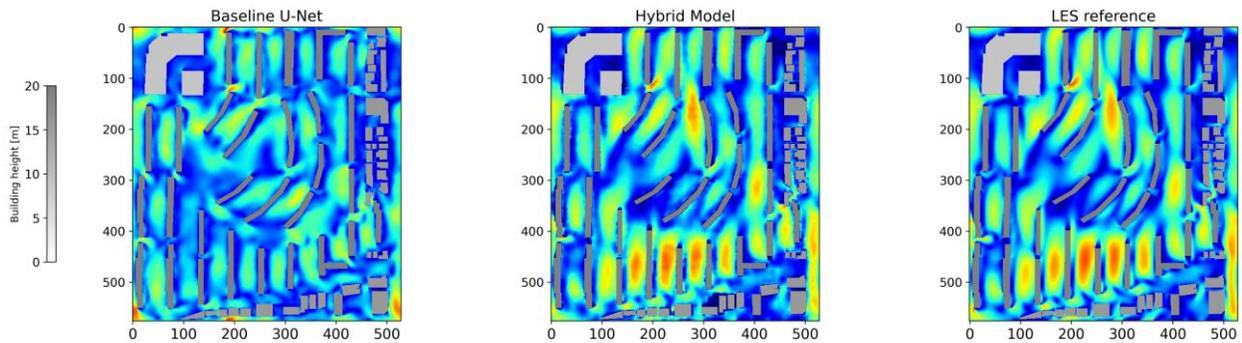

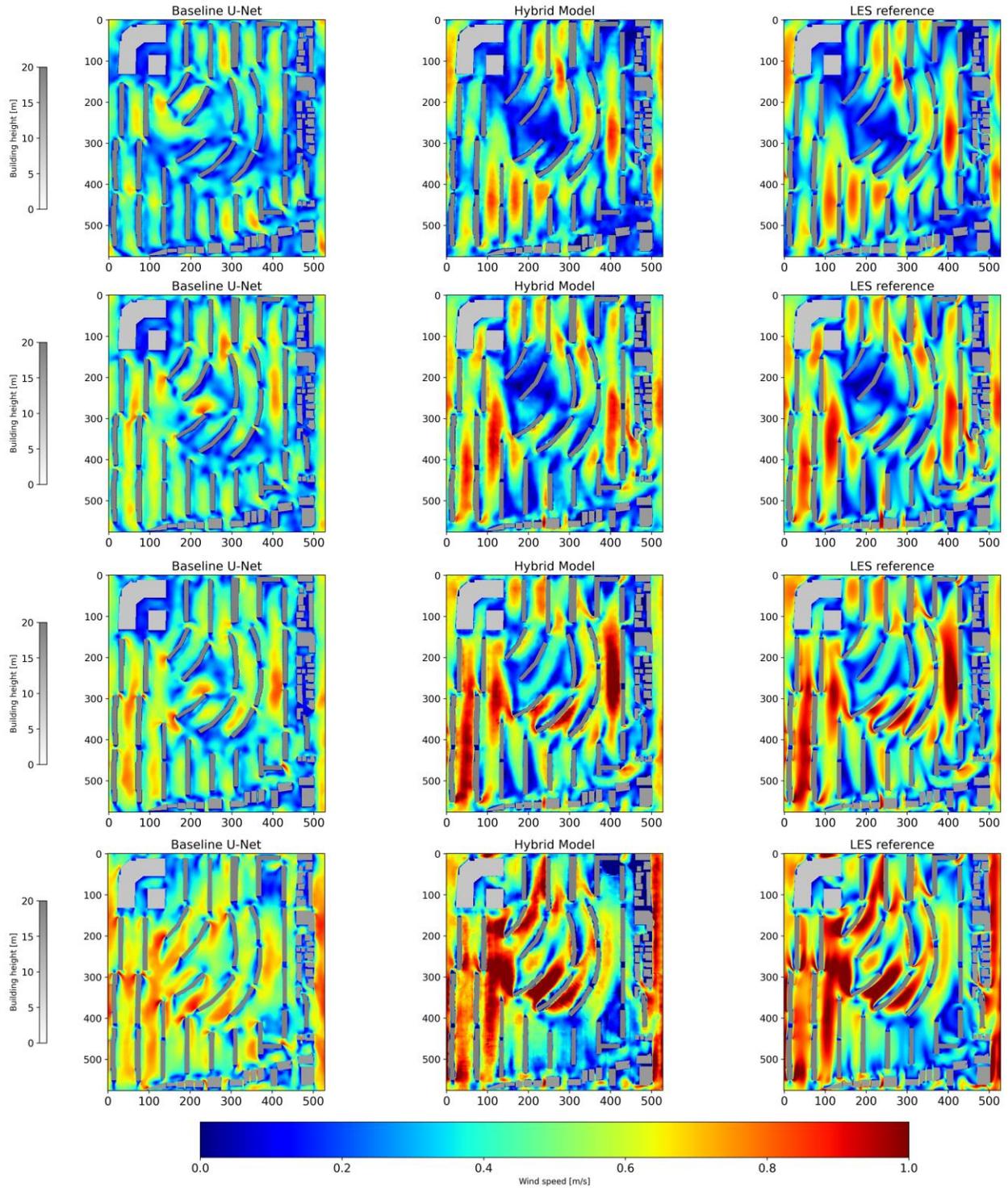

**Fig. 13**. Qualitative comparison of model performance under varying wind directions. Results for the KO-SE-V11 case under five different inflow angles: 0°, 15°, 30°, 45°, and 90°. Each row compares the baseline U-Net prediction (left), the final refined prediction (middle), and the LES ground truth (right).

As shown in Fig. 13, the refiner network improves the KO-SE-V11 predictions for all five inflow angles (0°, 15°, 30°, 45°, 90°). The qualitative gains are most evident for oblique (15°–45°) and perpendicular (90°) winds: the baseline captures broad recirculation but misses narrow high-speed jets and underestimates wake intensities, whereas the refiner restores concentrated jet streaks along passages and corrects wake magnitudes, yielding patterns that align closely with the LES fields. The quantitative results mirror these visual trends. RMSE decreases from 0.1607 to 0.0441 m/s at 0° (−72.6%), 0.2853 to 0.0521 m/s at 15° (−81.7%), 0.3698 to 0.0629 m/s at 30° (−83.0%), 0.3591 to 0.0615 m/s at 45° (−82.9%), and 0.2368 to 0.0847 m/s at 90° (−64.2%). Despite the strong angle dependence of the underlying flow physics, the refined model maintains a tight error band of 0.044–0.085 m/s across directions, indicating that the learned refinement is not tied to a single orientation and generalizes reliably to wind rotation.

## 4. Discussion

This study shows that decomposing the surrogate task into a two-stage predictor–refiner framework materially increases the fidelity of urban wind-field estimates. The Stage-1 U-Net delivers a fast, physically plausible baseline that preserves large-scale organization, while the Stage-2 conditional GAN (Pix2Pix with PatchGAN) specializes on the missing high-frequency content. The computational cost of this two-stage approach remains practical; on a standard CPU, the baseline U-Net requires approximately 0.639 s per scenario, with the refinement stage adding an overhead of 0.843 s for a total end-to-end latency of 1.482 s. This sub-two-second runtime is negligible compared to the LES simulations that provide the ground truth data. In exchange for this modest overhead, the refiner consistently recovers sharp shear zones, narrow jet streaks, and corrected wake intensities across diverse urban morphologies, features that are essential for pedestrian-level wind assessment.

Training dynamics support the plausibility of these gains. Using BCE-with-logits adversarial loss and an L1 reconstruction term with $\lambda = 100$, early stopping on validation L1 selected a stable checkpoint (best at epoch 146) with a small generalization gap (Train L1 0.007, Val L1 0.029, gap 0.022). Discriminator health metrics remained in a productive regime (median D-loss 0.325), and the generator's GAN-to-reconstruction contribution ratio at the selected checkpoint ($\approx$1.18) indicates that sharpening from the adversarial game complemented, rather than overwhelmed, pixel fidelity. Validation L1 varied by only ±0.004 around the minimum within a ±10-epoch window, suggesting convergence without late-epoch drift.

The quantitative evidence is consistent and strong. For six realistic cities, RMSE decreases by 50–82% on a per-case basis (FR-PA-V2: 66%, ES-Bar-V1: 66%, CA-Van-U2: 60%, BR-SP-U2: 82%, CH-Bas-V1: 55%, US-Det-U9: 51%), with corresponding tightening of dispersion across cities. On canonical regular grids, where the baseline is already comparatively well calibrated, the refiner still removes most residual error, yielding $\approx$84% RMSE reductions on average. Directional robustness is also evident: for KO-SE-V11, RMSE improves by 64–83% over 0°, 15°, 30°, 45°, 90°, while maintaining a tight absolute error band of 0.044–0.085 m/s across angles. Together, these regimes indicate that the learned mapping is not narrowly tuned to a specific morphology or orientation, but generalizes to the altered flow topologies induced by wind rotation and layout variability.

The architectural choices in the refiner appear central to these outcomes. Step-wise large receptive fields in the encoder (32→16→8) provide multi-scale context that helps correct low-frequency bias while preserving high-frequency detail via skip connections. Instance normalization and moderated dropout in early decoder layers balance stability and capacity. Crucially, training-set-only scaling to [−1,1] avoids leakage, and denormalization restores predictions to physical units before all metric computations, ensuring that reported RMSE values are interpretable in m/s.

While the model performs well across diverse cities and inflow angles, some limitations remain. The surrogate refines a single 2D plane (pedestrian height) of time-averaged wind speed, not the full 3D or unsteady flow; this enables tractable training and fast inference but omits vertical coupling (e.g., rooftop separation, lofting) and temporal variability relevant to comfort and safety. Only mean wind speed is supervised, therefore other key statistics of urban flow dynamics (turbulence intensity, TKE, gust factor, directional persistence, near-wall shear) are not guaranteed. In addition, the objective function combines adversarial and L1 terms without hard physics constraints (e.g., mass conservation or strict boundary adherence), so plausibility is learned statistically rather than enforced. Furthermore, training pairs come from a single corpus, so generalization to markedly different roughness, vegetation, thermal stratification, or mesoscale forcing is not assured without adaptation. Future work could address these points and others via multi-target supervision (adding TI/TKE/gusts), lightweight physics priors (divergence penalties, boundary consistency), uncertainty quantification, domain adaptation and active learning for high-error regimes, and extensions to 3D or time-aware models that retain the practical efficiency demonstrated here.

## 4. Conclusion

This study introduced a two-stage deep learning model for urban wind prediction that couples a U-Net predictor with a Pix2Pix cGAN-based refiner operating on the predictor's output. The two-step design preserves large-scale flow structure while the refiner restores high-frequency content missing from the baseline. Training converged stably: early stopping selected a best checkpoint at epoch 146 with a small generalization gap (Train MAE ≈ 0.010 m/s, Val MAE ≈ 0.041 m/s;

corresponding normalized L1s 0.007 and 0.029), and discriminator health metrics (median D-loss ≈ 0.33 with ~45% of epochs in a balanced range) indicated a productive adversarial game. Building on that foundation, the refiner consistently improved accuracy across representative evaluations: for realistic urban morphologies, RMSE decreased by ~50–80%; across inflow angles from 0° to 90° it decreased by ~73–83% with refined absolute errors clustered around 0.044–0.085 m/s; and on idealized grid arrays, reductions were ~83–85%. Qualitatively, the refiner sharpens gradients, recovers wakes and jets, and corrects magnitude biases without introducing artefacts, achieving high agreement with LES (typically $R^2 \geq 0.9$). Because training cost is far below LES and inference is near-instantaneous, the approach is well suited to early-stage urban design, pedestrian-wind comfort assessment, and rapid hazard screening.

**Declaration of Competing Interest**

The authors declare that they have no known competing financial interests or personal relationships that could have appeared to influence the work reported in this paper.

**Acknowledgements**

This work was partly supported by the Natural Sciences and Engineering Research Council of Canada (NSERC) [grant number CRSNG RGPIN 2022-03492].

**References**

BenMoshe, N., Fattal, E., Leitl, B. and Arav, Y., 2023. Using machine learning to predict wind flow in urban areas. *Atmosphere*, 14(6), p.990.

Blocken, B., 2015. Computational Fluid Dynamics for urban physics: Importance, scales, possibilities, limitations and ten tips and tricks towards accurate and reliable simulations. Building and environment, 91, pp.219-245.

Brunton, S.L., Noack, B.R. and Koumoutsakos, P., 2020. Machine learning for fluid mechanics. Annual review of fluid mechanics, 52(1), pp.477-508.

Buccolieri, R., Santiago, J.L. and Martilli, A., 2021, June. CFD modelling: The most useful tool for developing mesoscale urban canopy parameterizations. In *Building Simulation* (Vol. 14, No. 3, pp. 407-419). Beijing: Tsinghua University Press.

Chen, G., Wang, D., Wang, Q., Li, Y., Wang, X., Hang, J., Gao, P., Ou, C. and Wang, K., 2020. Scaled outdoor experimental studies of urban thermal environment in street canyon models with various aspect ratios and thermal storage. *Science of The Total Environment*, *726*, p.138147.

Clarke, A., Teigen Giljarhus, K.E., Oggiano, L., Saddington, A. and Depuru-Mohan, K., Deep Learning for Urban Wind Prediction: An Mlp-Mixer Approach with 3d Encoding. Available at SSRN 5225200.

Demir, U. and Unal, G., 2018. Patch-based image inpainting with generative adversarial networks. *arXiv preprint arXiv:1803.07422*.

Fenner, D., Christen, A., Grimmond, S., Meier, F., Morrison, W., Zeeman, M., Barlow, J., Birkmann, J., Blunn, L., Chrysoulakis, N. and Clements, M., 2024. Urbisphere-berlin campaign: investigating multiscale urban impacts on the atmospheric boundary layer. *Bulletin of the American Meteorological Society*, *105*(10), pp.E1929-E1961.

Goodfellow, I.J., Pouget-Abadie, J., Mirza, M., Xu, B., Warde-Farley, D., Ozair, S., Courville, A. and Bengio, Y., 2014. Generative adversarial nets. *Advances in neural information processing systems*, *27*.

Gråberg, H.M., 2022. *Towards Physics-Informed Neural Networks for Urban Wind Flow Prediction* (Master's thesis, NTNU).

Grimmond, C.S.B., King, T.S., Roth, M. and Oke, T.R., 1998. Aerodynamic roughness of urban areas derived from wind observations. *Boundary-Layer Meteorology*, 89(1), pp.1-24.

Isola, P., Zhu, J.Y., Zhou, T. and Efros, A.A., 2017. Image-to-image translation with conditional adversarial networks. In Proceedings of the IEEE conference on computer vision and pattern recognition (pp. 1125-1134).

Johnson, J., Alahi, A. and Fei-Fei, L., 2016, September. Perceptual losses for real-time style transfer and super-resolution. In European conference on computer vision (pp. 694-711). Cham: Springer International Publishing.

Kastner, P. and Dogan, T., 2023. A GAN-based surrogate model for instantaneous urban wind flow prediction. Building and Environment, 242, p.110384.

Li, Q., Chen, J. and Luo, X., 2024. Estimating omnidirectional urban vertical wind speed with direction-dependent building morphologies. *Energy and Buildings*, *303*, p.113749.

Lu, J., Nazarian, N., Hart, M.A., Krayenhoff, E.S. and Martilli, A., 2024. Representing the effects of building height variability on urban canopy flow. *Quarterly Journal of the Royal Meteorological Society*, *150*(758), pp.46-67.

Lu, J., Nazarian, N., Hart, M.A., Krayenhoff, E.S. and Martilli, A., 2023a. Novel geometric parameters for assessing flow over realistic versus idealized urban arrays. *Journal of Advances in Modeling Earth Systems*, *15*(7), p.e2022MS003287.

Lu, Y., Zhou, X.H., Xiao, H. and Li, Q., 2023b. Using machine learning to predict urban canopy flows for land surface modeling. Geophysical Research Letters, 50(1), p.e2022GL102313.

Lu, J., Li, W., Hobeichi, S., Azad, S.A. and Nazarian, N., 2025. Machine learning predicts pedestrian wind flow from urban morphology and prevailing wind direction. Environmental Research Letters, 20(5), p.054006.

Lyu, X., He, Y., Yin, S., Wong, S., Tse, T.K. and Ng, E., 2025. Evaluating urban design strategies for pedestrian-level ventilation improvement in a high-density urban living environment— A LiDAR and wind tunnel study. *Building and Environment*, 269, p.112439.

Mirza, M. and Osindero, S., 2014. Conditional generative adversarial nets. arXiv preprint arXiv:1411.1784.

Mittal, H., Sharma, A. and Gairola, A., 2018. A review on the study of urban wind at the pedestrian level around buildings. Journal of Building Engineering, 18, pp.154-163.

Mo, Z., Liu, C.H. and Ho, Y.K., 2021. Roughness sublayer flows over real urban morphology: A wind tunnel study. *Building and Environment*, 188, 107463.

Nazarian, N., Lipson, M. and Norford, L.K., 2023. Multiscale modeling techniques to document urban climate change. In *Urban climate change and Heat Islands* (pp. 123-164). Elsevier.

Nazarian, N., Lu, J., Lipson, M., Liu, S., Hart, M.A., Krayenhoff, S., Blunn, L. and Martilli, A., 2025. UrbanTALES: A comprehensive LES dataset for urban canopy layer turbulence analyses and parameterization development. eartharxiv.org

Ng, E., Yuan, C., Chen, L., Ren, C. and Fung, J.C., 2011. Improving the wind environment in high-density cities by understanding urban morphology and surface roughness: A study in Hong Kong. *Landscape and Urban planning*, *101*(1), pp.59-74.

OpenStreetMap contributors, 2017. Planet dump retrieved from https://planet.osm.org. https://www.openstreetmap.org.

Peng, W., Qin, S., Yang, S., Wang, J., Liu, X. and Wang, L.L., 2024. Fourier neural operator for real-time simulation of 3D dynamic urban microclimate. Building and Environment, 248, p.111063.

Shen, G., Zheng, S., Jiang, Y., Zhou, W. and Zhu, D., 2024. An improved method for calculating urban ground roughness considering the length and angle of upwind sector. Building and Environment, 266, p.112144.

Vargiemezis, T. and Gorlé, C., 2025. From large-eddy simulations to deep learning: A U-net model for fast urban canopy flow predictions. arXiv preprint arXiv:2507.06533.

Wu, T. and Snaiki, R., 2022. Applications of machine learning to wind engineering. Frontiers in Built Environment, 8, p.811460.

Xiao, D., Heaney, C.E., Mottet, L., Fang, F., Lin, W., Navon, I.M., Guo, Y., Matar, O.K., Robins, A.G. and Pain, C.C., 2019. A reduced order model for turbulent flows in the urban environment using machine learning. Building and Environment, 148, pp.323-337.

Xie, P., Li, T., Liu, J., Du, S., Yang, X. and Zhang, J., 2020. Urban flow prediction from spatiotemporal data using machine learning: A survey. Information Fusion, 59, pp.1-12.

Xu, J., Li, Z., Du, B., Zhang, M. and Liu, J., 2020. Reluplex made more practical: Leaky ReLU. In 2020 IEEE Symposium on Computers and communications (ISCC) (pp. 1-7). IEEE.

Yousif, M.Z., Yu, L. and Lim, H.C., 2022. Super-resolution reconstruction of turbulent flow fields at various Reynolds numbers based on generative adversarial networks. Physics of Fluids, 34(1).

Yu, Y., Li, P., Huang, D. and Sharma, A., 2024. Street-level temperature estimation using graph neural networks: Performance, feature embedding and interpretability. *Urban Climate*, *56*, p.102003.

Zhong, J., Liu, J., Zhao, Y., Niu, J. and Carmeliet, J., 2022. Recent advances in modeling turbulent wind flow at pedestrian-level in the built environment. *Architectural intelligence*, 1(1), p.5.